\documentstyle[12pt,psfig]{article}
\newcommand{\be}{\begin{equation}}
\newcommand{\ee}{\end{equation}}
\newcommand{\bea}{\begin{eqnarray}}
\newcommand{\eea}{\end{eqnarray}}
\newcommand{\norsl}{\normalsize\sl}
\newcommand{\norsc}{\normalsize\sc}
\begin{document}

\begin{titlepage}

\title{Perturbative QCD and Nucleon Structure Functions}
\author{
\norsc  Jiro KODAIRA\thanks{Electronic address:
kodaira@theo.phys.sci.hiroshima-u.ac.jp } \\
\norsl  Dept. of Physics, Hiroshima University\\
\norsl  Higashi-Hiroshima 724, JAPAN\\
}

\date{}
\maketitle

\begin{abstract}
{\normalsize
We review the basic aspects of the perturbative QCD based on the
operator product expansion to analyze the nucleon structure functions
in a pedagogical way.
We explain the non-trivial relation between the QCD results and
the parton model especially
to understand the polarized nucleon structure functions which deserve
much attentions in recent years.
}
\end{abstract}

\vspace{1.5cm}
\begin{center}
Invited talk at the\\
{\norsl
YITP Workshop on \lq\lq From Hadronic Matter to Quark Matter:\\
Evolving View of Hadronic Matter\rq\rq}\\
YITP, Kyoto Japan, October 1994\\
\end{center}
\vspace{0.5cm}
\begin{center}
{\norsl to be published in the proceedings}
\end{center}

\begin{picture}(5,2)(-300,-630)
\put(2.3,-110){HUPD\,-\, 9504}
\put(2.3,-125){January 1995}
\end{picture}

\thispagestyle{empty}
\end{titlepage}
\setcounter{page}{1}
\baselineskip 24pt

\section{Introduction}
\smallskip

The quantum chromodynamics (QCD) is a theory of the strong interaction.
So far all experimental data are consistent with the predictions
of QCD. Especially the high energy behavior of QCD is believed
to be described by the perturbation theory thanks to the asymptotically
free nature of QCD. Among many interesting and important problems,
the spin structure of nucleon has been one of the most exciting
subjects in recent years. The data on the polarized deep inelastic
scattering by the EMC collaboration~\cite{emc1} result in excitement
of not only particle physicists but also nuclear physicists since
the data seem to indicate that the nucleon's spin is not carried
by quarks (partons). This EMC experiment has incited many
(particle and nuclear) physicists to challenge
the so-called \lq\lq spin crisis\rq\rq\ problem in QCD.
After a flood of theoretical
papers as well as new experiments~\cite{emc2}-\cite{emc5},
our understanding on this problem is now much more improved:
the interpretation of
the QCD results in terms of the parton model is never obvious
and simpleminded one may fail: the axial anomaly
plays an important role: etc.

The aim of this talk is to provide a pedagogical introduction to the
perturbative QCD
to study the nucleon structure through the deep inelastic process
for non-experts of QCD who are interested in the
deep structure of the nucleon.
Those who are familiar with QCD and interested in recent
progress in this field are referred to recent nice
article~\cite{abook} and reviews~\cite{review}.

In Sect.2 we review the kinematics
of the deep inelastic lepton nucleon scattering process. In Sec.3
the basic approach of the perturbative QCD based on the operator product
expansion and the renormalization group equation will be explained.
The relation between the QCD results and the parton model
is discussed. It will be stressed that the parton
density is a \lq\lq conception\rq\rq\  which depends on the
renormalization
scheme. In Sec.4 we consider the polarized structure functions.
Concluding remarks including
some subtleties and/or controversial
aspects which deserve more investigations to understand recent
experimental data are given in Sec.5.

\bigskip
\section{The structure functions}
\smallskip

The cross section for the deep inelastic lepton ($l\,(k)$) nucleon
($N\,(p)$)
scattering $l\,(k) + N\,(p) \rightarrow l\,(k') + X $ (Fig.1)
\begin{figure}[h]
\begin{center}
\leavevmode\psfig{file=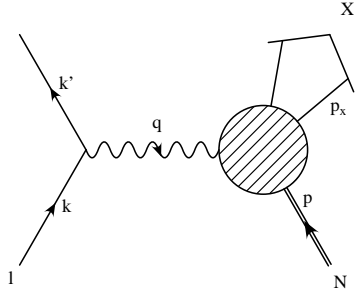,height=4cm}
\caption{Kinematic variables in inelastic lepton-nucleon scattering.}
\end{center}
\end{figure}
is given in terms of the leptonic and the hadronic tensors
according to the standard procedure in the field theory.
\[k'_0 \frac{d\sigma}{d^3k'} = \frac{1}{k\cdot p}
   \left( \frac{e^2}{4\pi Q^2}\right)^2 L^{\mu\nu}W_{\mu\nu}\ ,\]
where we consider only the QED interaction between the lepton and
nucleon
and keep only the lowest order in $\alpha_{QED}$. $q$ is the
momentum transfer
from the lepton to the nucleon and $q^2 \equiv -Q^2 = (k-k')^2$ .
The leptonic ($L_{\mu\nu}$) and the hadronic ($W_{\mu\nu}$) tensors
are defined as follows;
\[L_{\mu\nu} = \frac{1}{2} \sum_{s'} \langle
           k,s|j_{\mu}(0)|k',s'\rangle
                    \langle k',s'|j_{\nu}(0)|k,s \rangle\ , \]
\begin{eqnarray}
W_{\mu\nu} &=& \frac{1}{2 \pi}\sum_{X} \langle p,S|J_{\mu}(0)|X \rangle
               \langle X|J_{\nu}(0)|p,S \rangle (2\pi )^4 \delta ^4
               (p_{X} - p  - q) \nonumber \\
           &=& \frac{1}{2\pi }\int d^{\,4} x e^{iq\cdot x} \langle p,S|
               [J_{\mu}(x)\,,\,J_{\nu}(0)]|p,S \rangle \ .\label{ht}
\end{eqnarray}
with the lepton's (hadron's) electromagnetic current $j_{\mu}$
($J_{\nu}$). $s (S)$ is the spin of the lepton (nucleon).

In general, the spin 4-vector of the fermion with mass $m$ and
momentum $k$
is defined as ${\vec s\ }^2 = 1\,,\, s^2 = -1\,,\, s\cdot k = 0$.
So, for the longitudinally polarized (helicity $\pm$\ ) states, we
get,
\[s^{\mu}= \pm \frac{1}{m} (k\,,\,0\,,\,0\,,\,k^0)
                            \ \ ,\ \ k = |\vec{k}|^2 \ .\]
We can use the following approximation for leptons
$ms^{\mu}\simeq \pm k^{\mu}$ since $m_{\rm lepton} \simeq 0$.
Using this approximation, we get for the leptonic tensor,
\[
L^{\pm}_{\mu\nu}
           = k_{\mu}k'_{\nu}+k_{\nu}k'_{\mu}+\frac{q^2}{2}g_{\mu\nu}
           \mp i\varepsilon
        _{\mu\nu\lambda\sigma}q^{\lambda}k^{\sigma}\ .\]
On the other hand, the hadronic tensor contains all the information
of the strong interaction (QCD). Taking into account the various
symmetries, namely the Lorentz invariance, current conservation of
the QED current, T and P invariance,
we can write down the general form for $W_{\mu\nu}$,
\[W_{\mu\nu} \equiv W^{S}_{\mu\nu} + i\,W^{A}_{\mu\nu}\ ,\]
with
\begin{eqnarray*}
W^{S}_{\mu\nu} &=&
        -\,\left(\,g_{\mu\nu}-\frac{q_{\mu}q_{\nu}}{q^2}\right)
            W_1 + \left(\,p_{\mu}-\frac{p\cdot q}{q^2}q_{\mu}\right)
                   \left(\,p_{\nu}-\frac{p\cdot q}{q^2}q_{\nu}\right)
                   \frac{W_2}{M^2} \ ,\\
W^{A}_{\mu\nu} &=& \varepsilon _{\mu\nu\lambda\sigma}q^{\lambda}
              \left\{S^{\sigma}MG_1 + (p\cdot q S^{\sigma}-q\cdot S
                   p^{\sigma}) \frac{G_2}{M} \right\}\ .
\end{eqnarray*}
where $M$ is the mass of the nucleon. $W^{S}_{\mu\nu}$
($W^{A}_{\mu\nu}$) is the symmetric
(antisymmetric) part in $\mu\nu$ and relevant to the unpolarized
(polarized) process as shown below.
Usually we define the following dimensionless structure functions
(\,scaling functions\,):
\[F_1 \equiv W_1\ ,\ F_2 \equiv \frac{\nu}{2M}W_2\ ,\ g_1 \equiv
  \frac{M\nu}{2}G_1\ ,\ g_2 \equiv \frac{{\nu}^2}{2}G_2 . \]
with $M\nu \equiv p \cdot q$. These structure functions depend on the
$Q^2$ and $\nu$ or $Q^2$ and $x$
(Bjorken variable) $x \equiv \frac{1}{\omega} = \frac{Q^2}{2M\nu}$.

The explicit formula for the cross sections for this process
in the Laboratory frame corresponding to the configuration in Fig.2
is easily calculated to be~\cite{jaffe},
\begin{figure}[h]
\begin{center}
\leavevmode\psfig{file=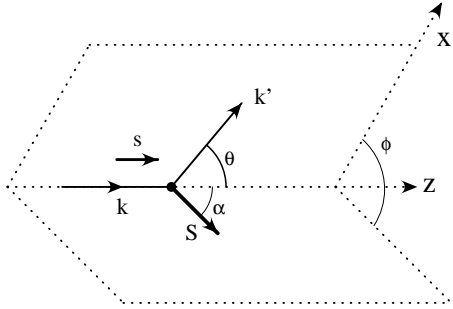,height=4cm}
\caption{Momentum and spin configuration in Lab. frame.}
\end{center}
\end{figure}
\[\frac{d \sigma^{\pm ,S}}{dE' d\Omega}
     = \frac{d\bar{\sigma}}{dE' d\Omega} \pm
              \frac{d \sigma^A }{dE' d\Omega} ,\]
with
\[\frac{d\bar{\sigma}}{dE' d\Omega} = \frac{2{\alpha}^2E^{\prime 2}}
    {Q^4 M} \left(\,2\,W_1 {\sin}^2 \frac{\theta}{2} + W_2 {\cos}^2
                \frac{\theta}{2} \right) ,\]
\begin{eqnarray*}
  \frac{d \sigma^A }{dE' d\Omega} &=&
     - \frac{\alpha^2 E'}{M Q^2 E}\,\,\Bigl[ \cos\alpha \left\{
      (E + E' \cos \theta )\,MG_1 - Q^2 G_2 \right\} \\
      & & \qquad\qquad + \sin\alpha\cos\phi \,\,E' \sin\theta
        \{ MG_1 + 2EG_2 \} \Bigr].
\end{eqnarray*}
The superscript $\pm$ refers to the lepton's helicity and $E'
\equiv k'_0$ and $E \equiv k_0$.
{}From these formulae, we can derive the expressions , for example, for
the longitudinal asymmetry which is the difference between the cross
section for the nucleon's spin being parallel
to the lepton's $(\uparrow\uparrow)$ and
the nucleon's spin being anti-parallel
to the lepton's $(\uparrow\downarrow)$ :
\[\frac{d{\sigma}^{\uparrow\downarrow}}{dE' d\Omega} -
    \frac{d{\sigma}^{\uparrow\uparrow}}{dE' d\Omega}
   = \frac{2{\alpha}^2E'}{M Q^2 E }
     \left\{MG_1(E + E' \cos \theta ) - Q^2 G_2 \right\} .\]

It is traditional and sometimes convenient to express the structure
functions in terms of the virtual photoabsorption cross sections
(in Lab. frame). The definition of the virtual photoabsorption cross
section is given by,
\[ \sigma_{\lambda} \equiv \frac{\pi e^2}{2M(\nu - Q^2 / 2M )}
  \epsilon^{\mu *}_{\lambda}(q) W_{\mu\nu}
       \epsilon^{\nu}_{\lambda}(q) ,\]
with the photon's polarization vector $\epsilon^{\mu}_{\lambda}$.
We have adopted the Hand-Berkelman's convention.
Taking the direction of photon's momentum $q$ to be z-axis
( note that $\vec{\mathstrut q} \neq \vec{\mathstrut k}$ ),
we have the following polarization vector for photons in the
nucleon's rest frame:
\[
   q^{\mu} = ( \nu,\, 0,\, 0,\, \sqrt{\nu^2 + Q^2} )\ ,\]
\[
 \epsilon_{R \atop L} = \frac{1}{\sqrt{2}}( 0,\, \mp i,\, 1,\, 0 )
     \,\,,\,\,
 \epsilon_S = \frac{1}{\sqrt{Q^2}}(\sqrt{\nu^2 + Q^2},\, 0,\, 0,\, \nu )
  \ .\]
For the unpolarized structure functions $W_1$ and $W_2$,
we get the relations:
\begin{eqnarray*}
  \sigma_S &=& \frac{\pi e^2}{2M(\nu - Q^2 / 2M )}
         \left[ W_2 \left( 1 + \frac{\nu^2}{Q^2} \right)
          - W_1 \right] ,\\
  \sigma_T &=& \frac{\pi e^2}{2M(\nu - Q^2 / 2M )} W_1 .
\end{eqnarray*}
where $T$ = $R$  and/or $L$.
The unpolarized lepton-nucleon scattering cross section is given
in terms of $\sigma_S$ and $\sigma_T$,
\[ \frac{d\bar{\sigma}}{dE' d\Omega} = \Gamma_T ( \sigma_T +
     \varepsilon \sigma_S)\ , \]
where
\[
\varepsilon^{-1} = 1 + 2\left( 1+ \frac{\nu^2}{Q^2} \right) \tan^2
            \frac{\theta}{2} \,\,\,,\,\,\,\,
\Gamma_T = \frac{\alpha}{2\pi^2} \frac{E'}{E} \frac{\nu - Q^2 / 2M}
          {Q^2 (1 - \varepsilon)}  .\]
Here we note that $\varepsilon$ means the ratio of $T$- and $S$-photon
present in the virtual photon .

To get the expressions for the polarized structure functions
$G_1$ and $G_2$, let us consider the three types of the photoabsorption
processes: $\sigma_{1/2}$ ($\epsilon_{L \atop R}$
with $S^{\mu}=(0,0,0,\pm 1)$); $\sigma_{3/2}$ ($\epsilon_{R \atop L}$
with $S^{\mu}=(0,0,0,\pm 1)$); $\sigma_{TS}$ (the interference
between $T$- and $S$- photon with $S_y = \pm 1$).
It is easy to obtain,
\begin{eqnarray*}
 \sigma_{1/2} &=& \frac{\pi e^2}{2M(\nu - Q^2 / 2M )}
          [ W_1 + M\nu G_1 - Q^2 G_2 ]\ ,\\
 \sigma_{3/2} &=& \frac{\pi e^2}{2M(\nu - Q^2 / 2M )}
          [ W_1 - M\nu G_1 + Q^2 G_2 ]\ ,\\
 \sigma_{TS}  &=& \frac{\pi e^2}{2M(\nu - Q^2 / 2M )}
          \sqrt{Q^2} [ M G_1 + \nu G_2 ]\ .
\end{eqnarray*}
Note that, $\sigma_T = \frac{1}{2} ( \sigma_{1/2} + \sigma_{3/2} )$ .
By defining the asymmetries $A_1$ and $A_2$:
\[
 A_1 \equiv
    \frac{\sigma_{1/2} - \sigma_{3/2}}{\sigma_{1/2} + \sigma_{3/2}}
    = \frac{M\nu G_1 - Q^2 G_2 }{W_1} \,\,,\,\,\,
 A_2 \equiv \frac{\sigma_{TS}}{\sigma_T} =
         \sqrt{Q^2} \,\,\frac{MG_1 + \nu G_2}{W_1} , \]
we can write the longitudinal asymmetry $A$ as;
\[
 A\,\, \equiv \frac{
      \frac{d{\sigma}^{\uparrow\downarrow}}{dE' d\Omega} -
      \frac{d{\sigma}^{\uparrow\uparrow}}{dE' d\Omega}}
     {\frac{d{\sigma}^{\uparrow\downarrow}}{dE' d\Omega} +
      \frac{d{\sigma}^{\uparrow\uparrow}}{dE' d\Omega}}
    = D ( A_1 + \eta A_2 ). \]
where
\[
   D = \frac{1 - (E'/E)\varepsilon}{1+ \varepsilon R}\,\,,\,\,\,
   \eta = \frac{\varepsilon \sqrt{Q^2}}{E - E' \varepsilon}\,\,,\,\,\,
   R \equiv \frac{\sigma_S}{\sigma_T} =
       \frac{\left(1+\frac{\nu^2}{Q^2}\right)W_2 - W_1}{W_1}.   \]

$D$ is called as the depolarization factor and its physical meaning
is obvious in Fig.3.
\begin{figure}[h]
\begin{center}
\leavevmode\psfig{file=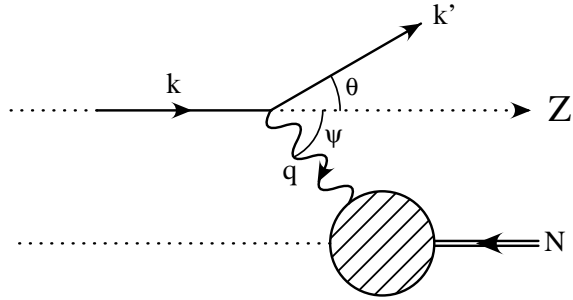,height=4cm}
\caption{Momentum configuration in inelastic lepton-nucleon
  scattering.}
\end{center}
\end{figure}
It is easily verified that,
\[D = D_1 \times D_2\ , \]
with
\begin{eqnarray*}
 D_1 &\equiv& \cos\psi = \frac{1 - (E'/E)\varepsilon}
                {\sqrt{1 - \varepsilon^2}} \ ,\\
 D_2 &\equiv& {\rm probability\,\, to \,\, have\,\,}T{\rm -photon}
              = \frac{\sqrt{1 - \varepsilon^2}\,\, \sigma_T }
                  {\sigma_T + \varepsilon \sigma_S }.
\end{eqnarray*}

The structure functions $g_1$ and $g_2$ can be expressed
in terms of $A_1$ and $A_2$.
\[
 g_1 = \frac{F_2}{2x ( 1 + R )}\,\left( A_1 +
           \frac{\sqrt{Q^2}}{\nu}\,A_2 \right)\,\,,\,\,\,
 g_2 = \frac{F_2}{2x ( 1 + R )}\,\left( \frac{\nu}{\sqrt{Q^2}}\, A_2
           - A_1 \right)\ ,
\]
where the following relation has been used,
\[ F_1 = W_1 = \frac{F_2}{x ( 1 + R )}\,\left( 1 + \frac{Q^2}{\nu^2}
           \right)\ . \]
Here it is to be noted that we have the
inequalities from the unitarity arguments ;
$\left| A_1 \right| < 1$ , $ \left| A_2 \right| < \sqrt{R} $.
Furthermore if we consider the scaling region (Bjorken limit),
$\frac{Q^2}{\nu^2} = \frac{4M^2 x^2}{Q^2} \ll 1$ ,
$g_1$ is given by,
\[ g_1 \cong \frac{A_1 F_2}{2x(1 + R)} .\]
This is the basic formula on which the experimental determination
of $g_1$ is based.

\bigskip
\section{The perturbative QCD}
\smallskip

In this section, we review the fundamental aspects of QCD to
analyze the structure functions introduced above. At first,
we discuss the general strategy of QCD based on the operator product
expansion (OPE) and the renormalization group equation (RGE). Next,
we consider the relation between the QCD results and the parton model
interpretations of the process we are considering.

\smallskip
\subsection{Formal approach in the perturbative QCD}
\smallskip

The kinematical region which we are interested in is the Bjorken
limit, namely $Q^2\,,\,\nu \rightarrow \infty $  with
$x = \frac{Q^2}{2M\nu}$ fixed. In this limit, we can easily recognize
that the hadronic tensor Eq.(\ref{ht}) is governed by the behavior
of the current products near the light-cone. So the light-cone
expansion, which is a variant of the OPE, of two currents might be
applied to this precess.
However the OPE can not be used directly here by the following reason.
The OPE makes sense in the short distance limit and this limit
corresponds to
the region where all component of $q_{\mu}$ become infinity. Therefore,
$Q^2 \geq 2p\cdot q$. On the other hand, the physical region for the
deep-inelastic process is $Q^2 \leq 2p\cdot q$.
Fortunately we can overcome this dilemma by using the dispersion
relation~\cite{chm} which relates the short distance limit to the
Bjorken limit for the deep-inelastic process.

Consider the time-ordered product of two currents which corresponds
to the forward Compton scattering of the virtual photon with
\lq\lq mass $q^2$ \rq\rq\  , (the Lorentz and the spin structure
being neglected),
\[T = i \int dx\,e^{iq\cdot x} \langle p|TJ(x)J(0)|
                   p \rangle ,\]
the physical region of which is
$2p\cdot q/|q^2| = 2M\nu / Q^2 \leq 1$. For this process, we can
use the OPE. In general, the product of two operators (currents)
can be expanded as follows;
\be
 TJ(x)J(0) \sim \sum_{i,n} C_i^n (x^2 -i\varepsilon )
       x^{\mu _1}x^{\mu _2}\cdots x^{\mu _n}
       O^{i}_{\mu _1 \cdots \mu _n}(0) ,\label{ope}
\ee
where $C_i^n$ is a c-number function called Wilson's coefficient
function and $O^{i}_{\mu _1 \mu _2 \cdots \mu _n}$
are local composite operators labeled by the index $i$.
The dimensional argument tells us that $C_i^n (x^2)$ behaves like,
\[C_i^n (x^2) \sim (x^2)^{(d_O^n - n - d_J - d_J)/2} ,\]
in the concerned limit of $x^2 \rightarrow 0$ with $x^{\mu}$ small but
$\neq 0$, where $d$ is the dimension of the corresponding operator.
So the operators with the lower twist ($\tau _N$):
\[\tau _N \equiv {\rm dim.} - {\rm spin} = d_O^n - n \]
dominate. The Fourier transform of Eq.(\ref{ope}) assumes the following
form with an appropriate normalization to define $C_i^n (Q^2)$,
\[i \int d^{\,4}xe^{iq\cdot x}TJ(x)J(0) \sim \sum_{i,n}
     C_i^n (Q^2)\left(\frac{2}{Q^2}\right)^n
     q^{\mu _1}\cdots q^{\mu _n}
     O^{i}_{\mu _1 \cdots \mu _n} .\]
By defining the matrix element of $O^{i}$ ,
\[\langle p|O^{i}_{\mu _1 \cdots \mu _n}(0)|p \rangle =
   2\,A^i_n p_{\mu _1}\cdots p_{\mu _n} - {\rm trace\ terms} ,\]
(note that the operators should have definite twists, so must be
traceless)
the forward Compton scattering amplitude $T$ is written as,
\[T(\nu ,Q^2) = 2\,\sum_{i,n} {\omega}^n A^i_n C_i^n (Q^2)\ \ ,\ \
          \omega = \frac{2M\nu}{Q^2} = \frac{1}{x} .\]

To get information for the structure functions in the Bjorken limit,
we rely on the analytic structure of $T$ in the complex $\omega$-plane.
There are cuts going out to infinity from $\omega = \pm 1$ and the
discontinuity of $T$ is related to the structure function $W$:
$W = \frac{1}{\pi}\,{\rm Im}T $. The next step is just to use
the Cauchy's theorem. The contour integral of $T$ around
the origin in $\omega$-plane picks up its n-th coefficient:
\[ \frac{1}{2\pi i}\oint d\omega {\omega}^{-n-1}T
     = 2\,\sum_i A^i_n C_i^n (Q^2) .\]
Deforming the contour to pick up the discontinuity of $T$,
the left hand side becomes,
\[ {\rm LHS} =
   \frac{2}{\pi}\int_1^{\infty}d\omega {\omega}^{-n-1}
              {\rm Im}T(\omega ,Q^2) = 2\int_0^1 dxx^{n-1}W(x,Q^2) ,\]
where we have used the crossing symmetry,
\[ T(q,p) = T(-q,p)\ ,\ W(q,p) = -\,W(-q,p) .\]
Finally we get the so-called moment sum rule,
\[\int_0^1 dxx^{n-1}W(x,Q^2) = \sum_i A^i_n C_i^n (Q^2) .\]
(In the above derivation, it is assumed that the integral is
convergent at infinity. Since this region corresponds
to the Regge limit,
the problem of convergence is controlled by the Regge behavior
of the corresponding amplitude.)

Now the Wilson's coefficient functions $C_i^n (Q^2)$ depend on only
the large momentum $Q^2$. So the RGE is effectively applied to
$C_i^n (Q^2)$ and they are calculated in the perturbation theory
due to the asymptotic freedom of QCD. On the other hand, the
hadronic matrix element of the composite operator $A^i_n$
can not be estimated perturbatively (long distance physics) and treated
as input parameters in the perturbative QCD.
The RGE for the coefficient functions is easily derived as follows.
Consider the Green's function of currents and
fundamental fields $\phi$;
\begin{eqnarray*}
  G_{JJ}^k &\equiv& \langle 0|TJ(x)J(0)\phi (x_1)
               \cdots \phi(x_k)|0 \rangle \\
            &=& \sum_n C^n (x)\langle 0|TO_n (0)\phi (x_1)
               \cdots \phi (x_k)|0 \rangle \equiv \sum_n C^n G_n^k ,
\end{eqnarray*}
where we write the OPE symbolically
as $J(x)J(0) = \sum_n C^n (x) O_n (0)$ .
The RGE for LHS reads,
\[\left[ {\cal D} + 2\,{\gamma}_J - k\gamma \right] G_{JJ}^k = 0 \,\]
where ${\gamma}_J \,,\,\gamma$\ the anomalous
dimensions of\ $J\,,\,\phi$ .
${\cal D}$ is the well-known operator in the obvious notation,
\[{\cal D} = \mu \frac{d}{d\mu} = \mu \frac{\partial}{\partial \mu}
      + \beta (g) \frac{\partial}{\partial g} - {\gamma}_m (g) m
        \frac{\partial}{\partial m}\ ,\]
with $m$ the quark mass. The RGE for RHS becomes,
\[\left[ {\cal D} + {\gamma}_O^n - k\gamma \right] G_n^k = 0 ,\]
with ${\gamma}_O^n$ the anomalous dimension of\ $O_n$\ .
\[{\gamma}_O^n = \mu \frac{d}{d\mu} \ln Z_O^n \quad , \quad O_n
          = \left( Z_O^n \right)^{-1}O_n^0 \ .\]
Therefore RGE for the coefficient functions is given by,
\[\left[{\cal D} + 2\,{\gamma}_J - {\gamma}_O^n \right] C^n = 0 .\]

It is easy to solve the RGE and the solution for the Fourier
transformed coefficient function becomes,
\begin{eqnarray*}
  C^{n}(Q^2) &=& C^{n}\left(\frac{Q^2}{\mu ^2}\,,\,g(\mu)\,,\,m(\mu)
          \right)  = C^{n}(\,1\,, \bar{g}(t)\,, \bar{m}(t)\,) \\
  & & \qquad \qquad \qquad \times \exp \int_0^t dt'
        \left[2\,{\gamma}_J (g(t')) - {\gamma}_{O}^n (g(t'))\right] ,
\end{eqnarray*}
where,
\[t = (1/2)\ln (Q^2 /{\mu}^2)\ ,\]
\[\frac{d \bar{g}}{dt} = \beta \left(\bar{g}\right)\ \ , \ \
  \frac{d \bar{m}}{dt} = -(\,1 + \gamma _m \left(\bar{g}\right))
           \bar{m}\ ,\]
with\ \ $\bar{g}(0) = g(\mu)\,,\,\bar{m}(0) = m(\mu)$. $\mu$ is
the renormalization point.
It is to be noted that the above solution takes, in general,
a matrix form because the operators with the same quantum number
and twist will mix under the renormalization.

Here let us consider how to calculate the anomalous dimension $\gamma$
and the \lq\lq coefficient function\rq\rq\ ,
$C^n (1, \bar{g}(t), \bar{m}(t))$.
The anomalous dimension $\gamma$ can be obtained from the
renormalization constant for the composite operator\ $O_n$.
To obtain $C^n (\,1,\bar{g},\bar{m})$ , we use the fact that the OPE
is an operator relation. So, consider a Green's
function which can be explicitly calculated, for example,
\begin{eqnarray*}
\Gamma (p,q) & = & i \int d^{\,4}xe^{iq\cdot x} \langle 0|T \phi(-p)
                    J(x)J(0)\phi ^{\dag}(p) |0 \rangle \\
             & = & \sum_n C^n (Q^2) \left(\frac{2}{Q^2}\right)^n
               q^{\mu _1}\cdots q^{\mu _n}\ \langle 0|T \phi(-p) O_n
                     \phi ^{\dag}(p) |0 \rangle .
\end{eqnarray*}
The LHS, the current correlation function, will be calculated to be,
\[\Gamma (p,q) = 2\,\sum_n t_n {\omega}^n\ \ ,\ \ t_n = t_n^0 + g^2
                   t_n^1 + \cdots .\]
The RHS becomes,
\[\Gamma (p,q) = 2\,\sum_n a_n C^n (Q^2){\omega}^n\ \ ,\ \ a_n =
                   a_n^0 + g^2 a_n^1 + \cdots ,\]
where $a_n$ is obtained from
\[\langle 0|T \phi(-p) O_n \phi ^{\dag}(p) |0 \rangle
     = 2 a_n p_{\mu _1}\cdots p_{\mu _n} - {\rm trace \ term} .\]
Therefore,
\be
    t_n = a_n C^n\ . \label{pc}
\ee
Expand $C^n (Q^2)$ in powers of $g$,
\begin{eqnarray*}
C^n (Q^2) & = & C^n (\,1,0\,) + g^2 \left. \frac{\partial C^n
           (\,1,\bar{g})}{\partial \bar{g}^2}\right|_{\bar{g}=0}\\
     & & \ \ - \frac{1}{2} g^2 {\gamma}^n_O \ln \frac{Q^2}{\mu ^2}
                    C^n (\,1,0\,) + {\cal O}(g^4 )\ ,
\end{eqnarray*}
where \ $\gamma ^n_O = {\gamma}^n_0 g^2 + {\cal O}(g^4)$ and,
for simplicity,
we neglect the mass $m$ and take into account the fact that
the anomalous dimension of the conserved current $J$ vanishes.
By comparing terms of the same power of $g$ in Eq.(\ref{pc}),
we will get $C^n (\,1,\bar{g})$.

Now the realistic case of QCD. The OPE of the electromagnetic currents
looks like,
\begin{eqnarray*}
\lefteqn{i\int d^{\,4}xe^{iq\cdot x}TJ_{\mu}(x)J_{\nu}(0)} \\
 & \sim & \left(g_{\mu\nu} - \frac{q_{\mu}q_{\nu}}{q^2}\right)
    \sum_{i,n} C_{L,i}^n (Q^2)\left(\frac{2}{Q^2}\right)^n
           q^{\mu _1}\cdots q^{\mu _n} O^i_{\mu _1 \cdots \mu _n}\\
   & & +\,(-g_{\mu\lambda}g_{\nu\sigma}q^2 + g_{\mu\lambda}
              q_{\nu}q_{\sigma} + g_{\nu\sigma}
              q_{\mu}q_{\lambda} - g_{\mu\nu}
              q_{\lambda}q_{\sigma}) \\
   & &  \qquad \qquad \qquad \qquad \times \sum_{i,n} C_{2,i}^n (Q^2)
              \left(\frac{2}{Q^2}\right)^n
                 q^{\mu _1}\cdots q^{\mu _{n-2}}
              O^i_{\lambda\sigma\mu _1 \cdots \mu _{n-2}} \\
   & & -i\varepsilon _{\mu\nu\lambda\sigma}q^{\lambda} \sum_{i,n}
              \left(\frac{2}{Q^2}\right)^n q^{\mu _1}\cdots
                         q^{\mu _{n-1}}\\
   & &  \qquad \qquad \qquad \qquad  \times
           [ E_{1,i}^n (Q^2) R^{1,i}_{\sigma\mu _1 \cdots \mu _{n-1}}
         + E_{2,i}^n (Q^2) R^{2,i}_{\sigma\mu _1 \cdots \mu _{n-1}}] .
\end{eqnarray*}
The explicit forms of the composite operators with the lowest twist
which contribute to the unpolarized structure
functions ~\cite{gw} read :
\begin{eqnarray*}
O^{i}_{\mu _1 \cdots \mu _n} & = & i^{n-1} S\bar{\psi}\gamma _{\mu _1}
          D_{\mu _2}\cdots D_{\mu _n}\frac{1}{2} \lambda ^i \psi\ , \\
O^{F}_{\mu _1 \cdots \mu _n} & = & i^{n-1} S\bar{\psi}\gamma _{\mu _1}
                      D_{\mu _2}\cdots D_{\mu _n} \psi\ , \\
O^G_{\mu _1 \cdots \mu _n} & = & \frac{1}{2}\, i^{n-2} S
         G_{\mu _1 \alpha}D_{\mu _2}\cdots D_{\mu _{n-1}}
                 G_{\mu _n}^{\alpha}\ ,
\end{eqnarray*}
where $\lambda ^i$ are the $SU(f)$ generators,
$D$\ is the gauge covariant derivative, and $S$ means the
symmetrization on the Lorentz indices.
To the spin dependent structure functions, the following twist-$2$
\,($R_1$) \cite{sar} and the twist-$3$\,($R_2$) \cite{sv}
operators contribute~\cite{hm}.
\bea
R^{1,i}_{\sigma\mu _1 \cdots \mu _{n-1}} & = & i^{n-1} S\bar{\psi}
         \gamma _5 \gamma _{\sigma}D_{\mu _1}
                      \cdots D_{\mu _{n-1}}\frac{1}{2} \lambda ^i \psi
                   \nonumber\ , \\
R^{1,F}_{\sigma\mu _1 \cdots \mu _{n-1}} & = & i^{n-1} S\bar{\psi}
         \gamma _5 \gamma _{\sigma}D_{\mu _1}\cdots D_{\mu _{n-1}} \psi
                  \label{g1op}\ , \\
R^{1,G}_{\sigma\mu _1 \cdots \mu _{n-1}} & = & \frac{1}{2}\, i^{n-1} S
      \varepsilon _{\sigma\alpha\beta\gamma}G^{\beta\gamma}D_{\mu _1}
       \cdots D_{\mu _{n-2}}G_{\mu _{n-1}}^{\alpha}\ , \nonumber
\eea
\bea
R^{2,F}_{\sigma\mu _1 \cdots \mu _{n-1}} & = & \frac{i^{n-1}}{n}
            \Bigl[ (n-1)\bar{\psi}
         \gamma _5 \gamma _{\sigma}D_{\{\mu _1}
                      \cdots D_{\mu _{n-1}\}}\psi \nonumber\\
          & & \quad - \sum_{l=1}^{n-1} \bar{\psi}
         \gamma _5 \gamma _{\mu_{l}}D_{\{\sigma}D_{\mu _1}
              \cdots D_{\mu _{l-1}}D_{\mu _{l+1}} \cdots
          D_{\mu _{n-1}\}}\psi \Bigr] \label{g2op}\ ,\\
R^{2,m}_{\sigma\mu _1 \cdots \mu _{n-1}} & = & i^{n-2} m \bar{\psi}
         \gamma _5 \gamma _{\sigma}D_{\{\mu _1} \cdots D_{\mu _{n-2}}
          \gamma _{\mu_{n-1}\}} \psi\ , \nonumber\\
R^{2,k}_{\sigma\mu _1 \cdots \mu _{n-1}} & = & \frac{1}{2n}
         ( V_k - V_{n-1-k} + U_k + U_{n-1-k} )\  ,\nonumber
\eea
where
\begin{eqnarray*}
 V_k &=& i^n g \bar{\psi}
         \gamma _5 D_{\mu _1} \cdots G_{\sigma \mu_{k}}
           \cdots D_{\mu _{n-2}}
          \gamma _{\mu_{n-1}} \psi \ ,\\
 U_k &=& i^{n-3} g \bar{\psi}
         D_{\mu _1} \cdots \widetilde{G}_{\sigma \mu_{k}}
           \cdots D_{\mu _{n-2}}
          \gamma _{\mu_{n-1}} \psi\ .
\end{eqnarray*}
We have shown only the flavor non-singlet operators for $R_2$.

Defining the matrix element of composite operators by,
\begin{eqnarray*}
\langle p,S|O^{j}_{\mu _1 \cdots \mu _n}|p,S \rangle & = & 2\,A^j_n
           p_{\mu _1}\cdots p_{\mu _n}\ , \\
\langle p,S|R^{1,j}_{\sigma\mu _1 \cdots \mu _{n-1}}|p,S \rangle
  & = & -2\,M a^j_n S_{\{\sigma}p_{\mu _1}\cdots p_{\mu _{n-1}\}}\, \\
\langle p,S|R^{2,F}_{\sigma\mu _1 \cdots \mu _{n-1}}|p,S \rangle
  & = & - 2\,M \frac{n-1}{n} d_n (S_{\sigma}p_{\mu _1}
          - S_{\mu _1}p_{\sigma})
           p_{\mu _2}\cdots p_{\mu _{n-1}}\ ,\\
\langle p,S|R^{2,m}_{\sigma\mu _1 \cdots \mu _{n-1}}|p,S \rangle
  & = & - 2\,M e_n (S_{\sigma}p_{\mu _1}
          - S_{\mu _1}p_{\sigma})
           p_{\mu _2}\cdots p_{\mu _{n-1}}\ ,\\
\langle p,S|R^{2,k}_{\sigma\mu _1 \cdots \mu _{n-1}}|p,S \rangle
  & = & - 2\,M f^k_n (S_{\sigma}p_{\mu _1}
          - S_{\mu _1}p_{\sigma})
           p_{\mu _2}\cdots p_{\mu _{n-1}}\ ,
\end{eqnarray*}
The moment sum rules are written as,
\begin{eqnarray*}
\int dx x^{n-1} F_L (x,Q^2) &=& \sum_j A^j_n C_{L,j}^n (Q^2) \ ,\
\int dx x^{n-2} F_2 (x,Q^2) = \sum_j A^j_n C_{2,j}^n (Q^2)\ , \\
\int dx x^{n-1} g_1 (x,Q^2) &=& \frac{1}{2}\, \sum_j
                a^j_n E_{1,j}^n (Q^2)\ , \\
\int dx x^{n-1} g_2 (x,Q^2) &=& -\frac{n-1}{2n}
     \Bigl[ a^i_n E_{1,i}^n (Q^2) - d_n E_{2,F}^n (Q^2) \Bigr]\\
      & & \qquad \qquad + \frac{1}{2} \Bigl[ e_n E_{2,m}^n (Q^2)
      + \sum_k f^k_n E_{2,k}^n (Q^2) \Bigr]\ .
\end{eqnarray*}
The each coefficient function ( $E^n$ or $C^n$ )
takes the following form neglecting the operator mixing,
\[E^n (Q^2) = E^n \left(1,\bar{g}(t)\right) \exp
           \left[- \int_0^t dt'\gamma ^n (\bar{g}(t'))\right] .\]
If we insert the perturbative results,
\bea
    E^n \left(1,\bar{g}(t)\right) & = & e^n_0 +
               e^n_1 {\bar{g}}^2 + \cdots \ ,\nonumber\\
   \gamma ^n (g) & = & \gamma ^n_0 g^2 + \gamma ^n_1 g^4 + \cdots
                       \label{pe} \ ,\\
   \beta (g) & = & -{\beta}_0 g^3 - \beta _1 g^5 + \cdots \nonumber,
\eea
we can get the final answer at the one-loop level,
\be
  E^n (Q^2) = N_n \left[e^n_0 + {\bar{g}}^2 \left\{
         e^n_1 + e^n_0 \left(\frac{\gamma ^n_1}{2\beta _0}
      - \frac{\beta _1 {\gamma}^n_0}{2\beta _0^2}\right)\right\}
       \right] {\bar{g}}^{\gamma ^n_0 / \beta _0} ,\label{rges}
\ee
where
\[N_n = g^{-\gamma ^n_0 /\beta _0}\left(1 + \frac{\beta _1}{\beta _0}
               g^2 \right)^{-\gamma ^n_1 /2\beta _1 +
              \gamma ^n_0 /2\beta _0} .\]
In this way, the $Q^2$ dependence of the structure functions can be
predicted in the perturbative QCD based on the OPE.

\smallskip
\subsection{Parton pictures in the perturbative QCD}
\smallskip

It is now a widespread belief that the results of the perturbative QCD
(at least at the lowest twist level) can be
understood in terms of the \lq\lq parton \rq\rq language.
In fact, the idea of the QCD improved
parton model~\cite{ap} has been justified for various processes
and has produced a great deal of progresses
in many hard reactions, especially those to
which we can not apply the OPE directly. As far as the deep
inelastic process, which we are interested in, is concerned, we do not
need the parton model at all since we can make a definite
prediction based only on the OPE and RGE. To relate the QCD results
based on the OPE with the parton picture, we must define
the parton distribution function in a appropriate way~\cite{kubk}.

Let us define the parton distribution function from the previous
results based on the formal approach of QCD.
Consider, for example, the moment of $F_2$\ in the singlet channel.
\[M_n \equiv \int dxx^{n-2} F_2 \ .\]
The QCD says,
\[M_n(Q^2) = A_n^F (Q_0^2)C^n_F (Q^2 /Q_0^2\,,\,g_0) +
             A_n^G (Q_0^2)C^n_G (Q^2 /Q_0^2\,,\,g_0)\ , \]
where
\[C^n_i (Q^2 /Q_0^2\,,\,g_0) =
         \left[T\exp \left\{-\int_0^t dt'{\gamma}^n
      (\bar{g}(t'))\right\}\right]^{ij} C^n_j (1,\bar{g}(t))\ , \]
and $A_n(Q^2_0)$ is the nucleon matrix element of the composite
operator renormalized at $Q^2_0$. The $Q^2_0$ dependence
is (should be) cancelled
between $A_n(Q^2_0)$ and the coefficient function.
Since $Q_0$ is arbitrary, let us put $Q_0^2 = Q^2$.
\[M_n (Q^2) = A_n^F (Q^2)C^n_F (1,\bar{g}(Q^2)) + A_n^G (Q^2)C^n_G
                      (1,\bar{g}(Q^2))\ .\]
Now we can define the \lq\lq {\it $Q^2$ dependent parton
distribution
function}\rq\rq\  of the quark and gluon;\ $q(x,Q^2)\,,\,g(x,Q^2)$ as,
\[A_n^F (Q^2) \equiv \int dxx^{n-1}q(x,Q^2)\ ,\ A_n^G (Q^2) \equiv
         \int dxx^{n-1}g(x,Q^2)\ . \]
Expanding $C^n$ perturbatively ,
\[C^n_F (1,\bar{g}) = 1 + c^n_{1,F}{\bar{g}}^2 + \cdots\ ,\ C^n_G
      (1,\bar{g}) = c^n_{1,G} {\bar{g}}^2 + \cdots \ ,\]
and making the inverse Mellin transformations, we get,
\begin{eqnarray*}
\frac{1}{x} F_2 (x,Q^2) & = & q(x,Q^2) + {\bar{g}}^2 \int_x^1
     \frac{dy}{y} K_F \left(\frac{x}{y}\right) q(y,Q^2) \\
              & & \ \ + {\bar{g}}^2 \int_x^1 \frac{dy}{y}
        K_G \left(\frac{x}{y}\right) g(y,Q^2) + {\cal O}({\bar{g}}^4) ,
\end{eqnarray*}
where
\[\int dxx^{n-1} K_F (x) = c^n_{1,F}\ ,\ \int dxx^{n-1} K_G (x) =
                  c^n_{1,G} .\]
We can interpret the above expressions as:
the first term describes the interaction of the charged parton (quark)
at the lowest order and other terms indicate the radiative corrections
for quarks and gluons at the higher orders.
So, the QCD results can be interpreted in terms of the parton model
with the $Q^2$ dependent parton distribution functions. Furthermore
we must take into account the presence of the radiative corrections.

It is here to be noticed that the above definition of the
parton distribution function (consequently, the corresponding
coefficient
functions too) has a degree of arbitrariness. Since the parton
distribution functions are defined as the nucleon matrix element
of the composite operator renormalized at $Q^2$ , they depend on how
to renormalize the composite operators. To understand the source
of the arbitrariness in the definition of parton distribution
functions, let us remember the procedure of
obtaining the coefficient functions.
In general, the quantities in Eq.(\ref{pc}) will be given as,
\begin{eqnarray*}
t_n & = & 1 + g^2 \left(d_n - \frac{1}{2} {\gamma}^n_0
                \ln \frac{Q^2}{-p^2}
            - {\gamma}_F \ln \frac{-p^2}{{\mu}^2} \right)
             + {\cal O}(g^4)\ , \\
a_n & = & 1 + g^2 \left\{b_n + \left(\frac{1}{2} {\gamma}^n_0 -
     {\gamma}_F \right) \ln \frac{-p^2}{{\mu}^2} \right\}
      + {\cal O}(g^4)\ , \\
C^n & = & 1 + g^2 \left(c^n_1 - \frac{1}{2} {\gamma}^n_0
               \ln \frac{Q^2}{{\mu}^2} \right) + {\cal O}(g^4)\ ,
\end{eqnarray*}
where we assume $t_n^0 = a_n^0 = c^n_0 = 1 $. ${\gamma}_F$ is the
anomalous dimension of the external field with which one estimates
the both sides of the OPE.
So, from Eq.(\ref{pc}) we get; $c^n_1 = d_n - b_n$.
Now $b_n$ depends on the renormalization scheme adopted.
Therefore $c^n_1$ also depends on the scheme.
The scheme ambiguity comes from how to renormalize the composite
operator. Suppose that the bare expression for $a_n$ to be,
\[\left( a_n \right)_{\rm bare} = 1 + g^2 \left\{\left({\gamma}_F
         - \frac{1}{2} {\gamma}^n_0 \right)
   \left( \frac{1}{\varepsilon} + \ln \frac{{\mu}^2}{-p^2} \right) +
      f_n \right\} + {\cal O}(g^4) \ .\]
To renormalize $a_n$, we multiply $Z_2 Z_{O_n}^{-1}$;
\[Z_2 = 1 - g^2 {\gamma}_F \frac{1}{\varepsilon} + {\cal O}(g^4)\ , \]
\[Z_{O_n}^{(j)} = 1 + g^2 \left( z_n^j - \frac{1}{2} {\gamma}^n_0
            \frac{1}{\varepsilon}\right) + {\cal O}(g^4) \ ,\]
where $j$ discriminate various renormalization schemes.
Therefore,
\[b_n = f_n - z_n^j\ , \]
and this means,
\[c^n_{1(j)} = d_n - (f_n - z_n^j) . \]
The ambiguity in $Z_2$ cancels
between $t_n$ and $a_n$, however $z_n^j$ is quite arbitrary.
So, the composite operators and also the coefficient functions
turn out to depend the renormalization scheme.

Physical quantities like moments should not depend on the
renormalization scheme adopted.
How does the above ambiguity cancel in the final expression ?
The difference between the schemes for the composite operator;
\[Z_{O_n}^{(j)} = \left[1 + g^2 ( z_n^j - z_n^k ) +
    {\cal O}(g^4)\right] Z_{O_n}^{(k)}\ , \]
affects the anomalous dimension of composite operator.
\begin{eqnarray*}
{\gamma}^{n(j)} & = & \mu \frac{d}{d\mu} \ln \left[1 + g^2
      ( z_n^j - z_n^k ) + {\cal O}(g^4)\right] + {\gamma}^{n(k)} \\
                 & = & -\,2 (z_n^j - z_n^k) {\beta}_0 g^4
                       + {\cal O}(g^6) + {\gamma}^{n(k)}\ .
\end {eqnarray*}
Then,
\[{\gamma}_1^{n(j)} = -\,2 (z_n^j - z_n^k) {\beta}_0 +
              {\gamma}_1^{n(k)}\ . \]
Therefore the ambiguity cancels in the final answer through the
combination,
\[c^n_1 + \frac{{\gamma}^n_1}{2\beta _0}\ \  .\]
in $C^n(Q^2)$. But each term can depend on the renormalization scheme.
We can say schematically,
\begin{eqnarray*}
 J J & \sim & \sum \left[ C^n_q O_n^q + C^n_G O_n^G \right] \\
     & = & \sum \left[ \widetilde{C}^n_q \widetilde{O}_n^q +
   \widetilde{C}^n_G \widetilde{O}_n^G \right]\ .
\end{eqnarray*}
LHS does not depend on the scheme.
However the arbitrariness remains in the definition of the parton
distribution functions.
The point is that the use of the parton language is helpful
to interpret the QCD results intuitively and/or economically.
So, the definition of parton distribution functions can
depend on one's convention.
It should be also mentioned that even if we start from the parton
model without using the OPE, the same ambiguities appear
at the stage of the factorization.

\bigskip
\section{The spin structure functions}
\smallskip

In this section, we will give a little bit more detailed analyses
for the spin structure functions $g_1$ and $g_2$. We will not mention
to the numerical studies of the recent experimental
data~\cite{review,alri,sloan}. Let us remember the moment sum rules
for $g_1$ and $g_2$ explained in sec.3~\cite{kmsu,k,kuy},
\bea
\int dx x^{n-1} g_1 (x,Q^2) &=& \frac{1}{2}\, \sum_j a^j_n
       E_{1,j}^n (Q^2) \label{g1}\ ,\\
\int dx x^{n-1} g_2 (x,Q^2) &=& -\frac{n-1}{2n}
            \Bigl[ a^i_n E_{1,i}^n (Q^2) - d_n E_{2,F}^n (Q^2) \Bigr]
                  \label{g2} \\
      & & \qquad \qquad + \frac{1}{2} \Bigl[ e_n E_{2,m}^n (Q^2)
      + \sum_k f^k_n E_{2,k}^n (Q^2) \Bigr] \nonumber.
\eea
The first moments $n=1$ of the above equations
are related to the interesting sum rules:
Bjorken sum rule~\cite{b}(the flavor nonsinglet part of $g_1$),
Ellis-Jaffe sum rule~\cite{ej} (the flavor singlet part of $g_1$)
and Burkhardt-Cottingham sum rule~\cite{bc} ($g_2$).

At first let us consider $g_2$. For general $n$,
the QCD analysis is not so straightforward for $g_2$ as for $g_1$.
The fact that the twist-3 operators also contribute to $g_2$ in the
leading order of $1/Q^2$ produces new aspects which do not appear
in the analyses of other structure functions.
The appearance of composite operators which are proportional
to the equation of motion makes the operator mixing problems
rather complicated. This is a general feature of the higher twist
operators. This problem has been discussed in Ref.\cite{kuy} and
references therein. However, as far as the first moment ($n=1$) of
$g_2$ is concerned, the above complexity is irrelevant since
there is no operators corresponding to $n=1$ (see Eq.(\ref{g2op})).
So, eq.(\ref{g2}) for $n=1$ predicts
the Burkhardt-Cottingham sum rule:
\[ \int dx g_2 (x,Q^2) = 0 . \]
Recently the validity of this sum rule was questioned~\cite{mn} and
explicit calculations of $g_2$ (current correlation functions)
at the one-loop level have been performed~\cite{alnr,kmsu2}.
The results confirm that the Burkhardt-Cottingham sum rule does
not receive any radiative corrections in the (at least)
perturbative QCD.

Next we turn to $g_1$ Eq.(\ref{g1}).
Three operators in Eq.(\ref{g1op}) contribute
to the moment of $g_1$. For convenience of explanation, let us
rewrite the operators and their proton's matrix elements
in the following way:
\bea
R^{1,i}_{n} & \equiv & i^{n-1} S\bar{\psi}_i
         \gamma _5 \gamma _{\sigma}D_{\mu _1}
                      \cdots D_{\mu _{n-1}} \psi _i \label{ns}\ , \\
R^{1,0}_{n} & \equiv & R^{1,F}_{\sigma\mu _1 \cdots \mu _{n-1}}
     \quad , \quad
R^{1,G}_{n} \equiv R^{1,G}_{\sigma\mu _1 \cdots \mu _{n-1}}\ ,
\label{s}
\eea
where $i$ is the the flavor index (we consider the case of
$3$ flavor $i$ = u, d, s):
\[
  \langle p,S|R^{1,j}_{n}|p,S \rangle \equiv  -2\,M
          \Delta ^j_n S_{\{\sigma}p_{\mu _1}\cdots p_{\mu _{n-1}\}}.
\]
Here $j$ runs over $i$, $0$ and $G$.
In this notation, the moment sum rule for $g_1$ Eq.(\ref{g1}) reads,
\bea
 I^n (Q^2 ) &\equiv & \int dx x^{n-1} g_1 (x,Q^2) \nonumber \\
           & = & \frac{1}{2}\, \Biggl[ \sum_i e_i^2 \Delta ^i_n
        E_{NS}^n (Q^2)  + \left\langle e^2 \right\rangle
              \left[ \Delta ^0_n E_{S}^n (Q^2)
        + \Delta ^G_n E_{G}^n (Q^2) \right] \Biggr] \label{g1new},
\eea
after taking into account the quark charge factor.
$\Delta ^0_n = \Delta ^{\rm u}_n +
\Delta ^{\rm d}_n + \Delta ^{\rm s}_n$ and $\left\langle e^2
\right\rangle = \sum_i e_i^2 /f = 2/9 $ for the number
of flavors being $f = 3$.
$E_{NS}^n \equiv E_{1,i}^n $ is the coefficient function for the
flavor non-singlet operator Eq.(\ref{ns}) and does not depend on $i$.
On the other hand $E_{S}^n \equiv E_{1,0}^n $
and $E_{G}^n \equiv E_{1,G}^n $ correspond
to the flavor singlet channel Eq.(\ref{s}) and get mixed under the
renormalization.

The QCD calculation of the coefficient functions at one loop level
has been done many years ago for both the
non-singlet~\cite{kmsu,kmmsu}
and singlet~\cite{k} parts. In the following, we restrict our
discussions only to the first ($n=1$) moment.
For a nucleon target (p: proton, n: neutron), the first moment
becomes from Eq.(\ref{g1new}),
\be
 I^1_{p,n} (Q^2 ) = \frac{1}{12}\, \Biggl[ E_{NS}^1 (Q^2)
   \left[ (+-)(\Delta ^{\rm u}_1 - \Delta ^{\rm d}_1 ) + \frac{1}{3}
             (\Delta ^{\rm u}_1 + \Delta ^{\rm d}_1
       - 2 \Delta ^{\rm s}_1) \right] + \frac{4}{3} E_{S}^1 (Q^2)
           \Delta ^0_1 \Biggr] \label{g11},
\ee
Since the gluon operator (corresponding coefficient function)
with $n=1$ does not exist (see Eq.(\ref{g1op})),
only the fermion bilinear operators contribute which
turn out to be the axial vector currents.
Note that the anomalous dimension of the non-singlet axial current
vanishes because of the current conservation in (massless) QCD.
Therefore the nucleon's matrix elements of the non-singlet
axial currents are scale independent and given by,
\bea
  \Delta ^{\rm u}_1 - \Delta ^{\rm d}_1 =
        \frac{G_A}{G_V} &\cong  & 1.26 \ ,\nonumber\\
  \Delta ^{\rm u}_1 + \Delta ^{\rm d}_1 -
         2 \Delta ^{\rm s}_1 &\cong & 0.58 \ ,\nonumber
\eea
using the flavor SU(3) symmetry.
The corresponding coefficient function has been calculated in
the perturbative QCD up to the three loop in $\overline{{\rm MS}}$
scheme~\cite{ltv},
\[
  E_{NS}^1 (Q^2) = 1 - \frac{\alpha _S }{\pi}
       - 3.58 \left( \frac{\alpha _S }{\pi}\right)^2
       - 20.2 \left( \frac{\alpha _S }{\pi}\right)^3 + \cdots ,\]
where $\alpha _S \equiv \bar{g}^2 (Q^2 )/4\pi$.
Since the Bjorken sum rule receives the contribution only from the
non-singlet
channel, we can make a definite QCD prediction for this sum rule
in the sense that the matrix element is known.
\[ I^1_p - I^1_n = \frac{1}{6}\frac{G_A}{G_V}
         \left[ 1 - \frac{\alpha _S }{\pi}
   - 3.58 \left( \frac{\alpha _S }{\pi}\right)^2
   - 20.2 \left( \frac{\alpha _S }{\pi}\right)^3 + \cdots \right].\]

The flavor singlet axial current is not conserved due to the
Adler-Bell-Jackiw anomaly~\cite{abj}. This fact makes the matrix
element scale dependent~\cite{k,j}. The perturbative result
for $E^1_S(Q^2)$ is given as (see Eqs.(\ref{pe},\ref{rges})),
\[
  E^1_S(1,\bar{g}(t)) = 1 - \frac{\alpha _S }{\pi} + \cdots .\]
Due to the anomaly, the anomalous dimension $\gamma$ starts at the
two loop~\cite{k},
\[ \gamma ^1_0 = 0 \quad , \quad \gamma ^1_1 =
           \frac{1}{(16\pi ^2 )^2}24 C_2 (R) T(R).\]
(For the higher order corrections, see Refs.\cite{znl}.)
Using these results, we write the singlet parts as follows.
At first, let us define $\Delta\Sigma (Q^2)$ by (see section 3.2),
\[ E_{S}^1 (Q^2) \Delta ^0_1 (\mu)
      \equiv E^1_S(1,\bar{g}(t)) \Delta\Sigma (Q^2) .\]
Then,
\bea
  \Delta\Sigma (Q^2) &=& exp \left[ - \int_0^t dt'
    \gamma ^1(\bar{g}(t')) \right] \Delta\Sigma (\mu ^2)\nonumber\\
      &=& \left( 1 + \frac{\alpha _S }{\pi}
    \frac{6f}{33-2f} + \cdots \right) \Delta\Sigma (\infty)\nonumber
\eea
The final expression becomes,
\[ E^1_S(1,\bar{g}(t)) \Delta\Sigma (Q^2)
    = \left( 1 - \frac{\alpha _S }{\pi}
       \frac{33-8f}{33-2f} + \cdots \right) \Delta\Sigma (\infty).\]

In the remainder of this section, we will mention to the
so called \lq\lq spin crisis\rq\rq or
\lq\lq spin deficit\rq\rq~\cite{sloan}
problem. At first we want to stress that this is never the problem
of QCD. From the viewpoint of the formal approach, we have already
finished all tasks to predict the moment
of the structure functions. What we should do next is just to
compare the above results with the experimental data.
On the other hand, it is also true that the interpretation of QCD
in terms of the parton language is very helpful and convenient
to understand the hard processes. As explained in section 3.2,
the definition of parton densities depends on the scheme; then
is ambiguous.
In most cases like the unpolarized structure functions, however,
this ambiguity produces only small differences and a naive
parton model interpretation holds rather well (of course, with
the $Q^2$ parton densities). If it is the case also for $g_1$,
we will have,
\[ \Delta\Sigma  = \sum_i \int_0^1 dx \Delta q_i(x) ,\]
where
\[ \Delta q_i(x) \equiv q_{i+}(x) + \bar{q}_{i+}(x)
                 - q_{i-}(x) - \bar{q}_{i-}(x) ,\]
with $q_{i \pm}$ ($\bar{q}_{i \pm}$) being the density of
flavor i quark (antiquark)
parton with helicities $\pm 1/2$. So if the nucleon's spin
is carried by charged partons, we will expect
$\Delta\Sigma \sim 1$. The first EMC result~\cite{emc1}
was $\Delta\Sigma \sim 0.1$ ($\Delta^{\rm s}_1 \sim - 0.2$)!!
(for the recent status of data, see Ref.\cite{alri}.)

A resolution to this problem was proposed in Ref.\cite{ar}.
In the approach of the QCD improved parton model, they have derived
the relation,
\[ \Delta\Sigma (Q^2) = \sum_i \int_0^1 dx \left( \Delta q_i(x,Q^2)
       - f\frac{\alpha_S}{2\pi} \Delta g(x,Q^2) \right) ,\]
with
\[ \Delta g(x,Q^2) \equiv g_{+}(x,Q^2) - g_{-}(x,Q^2) .\]
$g_{\pm}$ is the gluon density with helicities $\pm 1$.
This result says that $\Delta\Sigma$ does not measure
the spin fraction carried by quarks.
Furthermore, it was also shown that
the second term is independent of the scale $Q^2$ in the leading
logarithmic approximation. (There may be some nonperturbative
corrections to the above relation.
See Ref.\cite{review} for the details.)
So if the second term assumes some finite value, a smallness of
$\Delta\Sigma$ is not necessarily in contradiction with our naive
expectation.
However the story is not so simple because we can reach a different
conclusion even if we work in the the QCD improved parton model
depending on the regularization of soft singularities which
turn out to be the scheme dependence~\cite{l}.

Now how does the above result reconcile with the approach
based on the OPE ? We do not have a gauge invariant operator
composed from the gluon field which may correspond to $\Delta g$.
The answer is the following:
we just introduce the \lq\lq Chern-Simons current\rq\rq\ $k_{\sigma}$
as $n=1$ gluonic operator,
\[k_{\sigma} = {\varepsilon}^{\sigma\mu\nu\lambda} A_{\mu}^a
                   ( {\partial}_{\nu} A_{\lambda}^a -
                     \mbox{$\frac{1}{3}$} g f_{abc} A_{\nu}^b
                                 A_{\lambda}^c ) .\]
In this case, we can make a finite renormalization of the composite
operators between the axial current $R_1^{1,0}$ and $k_{\sigma}$.
And we can get under a particular scheme that the coefficient
function $E_G^1$ which is associated with $k_{\sigma}$ takes
the value,
\[E_G^1 (1,\bar{g}) = -\frac{\alpha_S}{\pi} T(R)
                            = -\frac{\alpha_S}{\pi}\frac{f}{2} .\]
Therefore,
\[  \Delta\Sigma (Q^2) \sim  \Delta^q
                     - f \frac{\alpha_S}{2\pi} \Delta^G , \]
with,
\[
\langle P,S| \left[ R_1^{1,0} \right]_R |P,S \rangle
             =  -2M \Delta^q S_{\sigma} \ ,\
\langle P,S| \left[ k_{\sigma}\right]_R |P,S \rangle
             =  -2M \Delta^G S_{\sigma} ,\]
where the subscript $R$ denotes the renormalized operator.
If we identify $\Delta^q$ ($\Delta^G$) to be the moment of
$\sum_i \Delta q_i(x)$ ($\Delta g (x)$), there is no contradiction.
In the OPE of the electromagnetic currents, only the gauge (BRS)
invariant operators can appear. The Chern-Simons current, however,
is gauge dependent. This gap will be filled, if one notes
the following trick.
In the OPE, only the operator $R_1^{1,0}$ in fact appears. But we can
always write as,
\[  R_1^{1,0} = \left( R_1^{1,0} + f \frac{g^2}{8\pi ^2}
         k_{\sigma}\right)
                      - f \frac{g^2}{8\pi ^2}k_{\sigma} .\]
The first (second) term will be identified as quark (gluon) densities.
In fact, in the above renormalization scheme,
\[  \left[ R_1^{1,0} \right]_R = R_1^{1,0} +
               f \frac{g^2}{8\pi ^2}k_{\sigma} ,\]
at the lowest order.

Now ambiguities can come in the above game. Even if we introduce
$k_{\sigma}$,
we can get different answer depending on how one
renormalizes the composite operators. Namely we can put an
arbitrary number in the front of the Chern-Simons current by
choosing a different scheme. This fact exactly
corresponds to the ambiguities which appear also
in the QCD improved parton model approach mentioned before.
In this way, the definition of the parton densities inevitably
depends on the scheme. So it is a waste of time to make a discussion
without specifying the scheme one adopted.
The point is :
Which scheme will be convenient to parametrize the parton
densities and easy to understand the physics ?
And once one defines the parton distribution functions,
whether one can explain other processes ?
(The definition in Ref.\cite{ar} seems to be reasonable
in the sense that the quark helicity $\Delta_q$ is conserved
to all orders since the corresponding operator is conserved
current by construction.)

\bigskip
\section{Concluding Remarks}
\smallskip

We have explained, in this review, only the first stage of
the approach in the perturbative QCD to analyze the structure
of nucleons.
New experiments at CERN~\cite{emc2,emc3} and SLAC~\cite{emc4,emc5}
and detailed theoretical investigations
have enormously improved our understanding on the (spin) structure
of nucleon. Now all data seem to be consistent with
the expectations from the perturbative QCD. It is impressive that
we can make a precision test of QCD to the accuracy of
about 10\%~\cite{alri,ek} for the spin structure of nucleons.
Furthermore, the problem of \lq\lq spin crisis\rq\rq\  became
less exciting. We can construct a consistent picture in terms
of the parton language at least qualitatively
for the spin structure function~\cite{review,y}.

On the other hand, we would like to emphasize that
there still remain many subtleties and controversial aspects
when one tries to make a much more precisely quantitative prediction
of QCD. We will list up below several points
which will need more investigations.
To compare the experimental data with the QCD prediction,
we must take into account many theoretical corrections
and sometimes make several assumptions.
Among those problems are : the small $x$ (and large $x$ )
behavior of the structure
functions~\cite{sx}: theoretical treatments of the \lq\lq wee\rq\rq\
partons including the role of the s quark
and the iso-spin SU(2) violation~\cite{wee}: the $Q^2$ dependence
of the structure
functions both in theories and experiments~\cite{anr}: etc.
Since the experimental data on $g_1$ are taken at rather small
values of $Q^2$, the power corrections in $Q^2$ to the sum rule
may give some modifications. The power corrections come from
the higher twist terms as well as the
target mass effects~\cite{ku}.
For the higher twist effects, many authors have
tried to reveal their theoretical behavior and numerical
importance in the analyses of data~\cite{ht}. However
it has been pointed out recently
that there is not only phenomenological but also
theoretical problem in including the higher twist
contributions~\cite{m}.
Due to the renormalon singularities, there appears some degree
of arbitrariness in the twist expansion.
These problems require much more elaborate theoretical investigations.
Forthcoming precision measurements will also provide us with
much information and clues to answer the above questions.

As explained in the text, parton densities depend on the scheme.
Therefore to confirm the parton model interpretation
of the spin structure function, we need independent experiments
which could measure the parton (quark and gluon) densities.
If we consider a quite different process (e.g. Drell-Yan process),
we encounter other spin-dependent structure functions.
The experimental and theoretical studies~\cite{yk} of these
structure functions must help us to fully understand the
spin structure of nucleons.
Nuclear effects in measuring the neutron structure functions
and nuclear structure functions are also interesting
and important. We need more
investigations~\cite{thom,ak} on these subjects.

Finally, we hope that various kinds of new experiments
and more detailed theoretical investigations will be able to
clarify the not only the perturbative but also the nonperturbative
aspects of QCD related to the Spin Physics.

%
\vspace{3cm}
\noindent
{\large\bf Acknowledgements}
\medskip

\noindent
The author is indebted to S.Matsuda, K.Sasaki, T.Uematsu and Y.Yasui
for invaluable discussions and for fruitful collaborations on various
topics discussed here. He also thanks H.Kawamura and
T.Nasuno for useful comments in
preparing this talk. This work is supported in part by the Monbusho
Grant-in-Aid for Scientific Research No. C-05640351.


%
\end{document}